\documentstyle[aps,pra,psfig,epsfig,twocolumn]{revtex}

\begin{document}
\title{Electron (hole) paramagnetic resonance of spherical CdSe nanocrystals.}
\author{K. Gokhberg$^{1}$, A. Glozman$^{1}$, E. Lifshitz$^{1}$ , T. Maniv$^{1}$ ,
M.C. Schlamp$^{2}$ and P. Alivisatos$^{2}$}
\address{$^{1}$Department of Chemistry and Solid State\\
Institute,Technion--Israel Institute of Technology, \\
Haifa 32000, Israel\\
$^{2}$Department of Chemistry, University of California\\
and Material Science Division, \\
Lawrence Berkeley National Laboratory,\\
Berkeley, Cal. 94720, USA}
\maketitle

\begin{abstract}
A new mechanism of electron paramagnetic resonance in spherical zinc-blende
semiconductor nanocrystals, based on the extended orbital motion of
electrons in the entire nanocrystal, is presented. Quantum confinement plays
a crucial role in making the resonance signal observable. The mechanism
remains operative in nanocrystals with uniaxially distorted shape. A
theoretical model based on the proposed mechanism is in good quantitative
agreement with unusual ODMR spectra observed in nearly spherical CdSe
nanocrystals.
\end{abstract}

Nearly two decades of research have explored the origin and potential
applications of unusual opto-electronic properties of semiconducting
nanocrystals. In particular, quantum size effects on nanometer length scale
can lead to unique features in optical properties (\cite{Bawendi},\cite
{Eychmuller}) as compared to the corresponding properties of the bulk
solids. Several reports have recently described the development of a new
type of colloidal nanocrystals based on CdSe and CdS core coated with an
epitaxial shell of a different semiconductor or by organic ligands (\cite
{Alivisatos}-\cite{Hasselbarth}).

The electronic properties of nanocrystals have been investigated extensively
during the last decade (see e.g. \cite{Glozman},\cite{Efros2},\cite{Norris}
and references therein). However, there is still significant uncertainty
regarding the orbital and spin dynamics of carriers in the presence of an
external magnetic field, under the quantum confined conditions of the
nanocrystals. In particular, a direct method for determining the carriers'
effective masses in nanocrystals, similar to the cyclotron resonance
technique used routinely in bulk semiconductors, is lacking. \ As will be
shown in this paper, however, application of the Optically Detected Magnetic
Resonance (ODMR) technique to nanocrystalline materials can provide the
missing information by taking advantage of the quantum confinement imposed
on the carriers dynamics.

Our recent ODMR study of CdSe and CdS nanocrystals has revealed unusually
broad resonance bands which could not be accounted for in terms of the
standard spin Hamiltonian model since such an interpretation would yield
unreasonably large g-factors ( i.e. in the range of $4-10$ ). \ In this
paper we present a new mechanism of electron paramagnetic resonance in
spherical semiconducting nanocrystals, based on the extended orbital motion
of electrons in the entire nanocrystal, which can account for the observed
resonance bands. \ In macroscopic crystals paramagnetic resonance of
conduction electrons is strongly damped by the coupling to the lattice
degrees of freedom ( typical spin-lattice relaxation times $T_{1}$ are in
the range of $10^{-10}$ to $10^{-8}$ sec. (\cite{PikTit})). The dramatic
suppression of the relaxation rates in small nanocrystals ( with $T_{1}\sim
10^{-6}-10^{-4}$ sec\ (\cite{Glozman}) ) due to the effect of quantum
confinement can make these resonances easily observable.

Nanocrystals of CdSe, capped either with tri-octy-phosphine-oxide (TOPO) or
epitaxial layers of CdS with core diameter of about 30$\AA $, were prepared
according to the procedure described by Peng et al.~(\cite{Peng}). The
photoluminescence (PL) and ODMR measurements were carried out by immersing
the samples in a cryogenic dewar (at 1.4K) and exciting them with a
continuous 457.9 nm $Ar^{+}$ laser. The sample was mounted on a special
sample probe at the centre of a High-Q resonance cavity, coupled to a
microwave (mw) source (~10 GHz), and surrounded by a superconducting magnet
(B). A detailed description of the experimental system is described in Ref.( 
\cite{Glozman}). The ODMR spectra were obtained by measuring a change in
luminescence intensity, due to the absorption of a modulated magnetic
component of a microwave radiation. The latter was plotted (see Fig.1)
versus the external magnetic field, B, yielding a magnetic resonance-like
spectrum.

The PL spectra of CdSe (TOPO) and CdSe/CdS core-shell samples consist of an
exciton band (centred at 2.15 eV), predominantly tunable with the size of
the core, and an additional broad band at lower energies (centered at 1.70
eV). The exciton band in both samples did not show resonance phenomena and
therefore, the ODMR spectra were selectively recorded around 1.7 eV. The
ODMR spectrum of either CdSe(TOPO) or CdSe/CdS nanocrystals consisted of
three broad resonance bands (Fig. 1a), ranging between 200-8500 Gauss. The
spectrum of the CdSe/CdS core-shell structure showed additional overlapping
narrow resonance bands (see inset in Fig. 1a ) between 3000-5000 Gauss. The
broad resonances were mainly pronounced upon an application of kHz mw power
modulation, while the narrow resonances in CdSe/CdS were observed under a mw
modulation of about 100Hz. This distinction suggests the existence of two
types of magnetic resonance events, with different characteristic relaxation
times. Furthermore, the appearance of the narrow resonances with g-factors
close to 2.000 ( i.e. $1.988$ and $1.845$ ) in contrast to the broad ones,
emphasizes the localized nature of the corresponding states within the
forbidden band gap. The chemical identification of the corresponding
trapping sites are discussed at length in a separate publication ~(\cite
{Glozman}). They are considered as pure spin resonance transitions of
trapped electrons and holes.

\begin{figure}[htb]
\begin{center}
\psfig{file=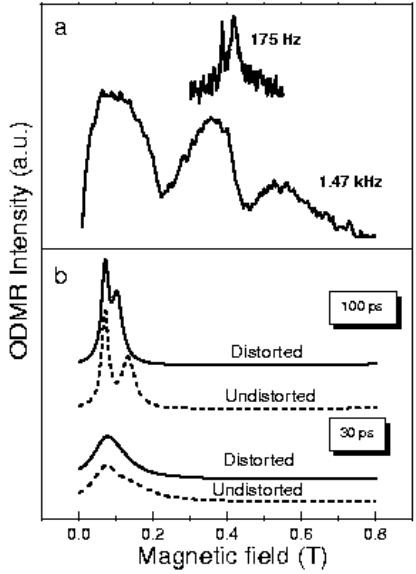,width=2.5in}
\caption{(a) ODMR spectra of CdSe(TOPO) recorded at 1.47 kHz. The inset
corresponds to the ODMR spectrum of CdSe(CdS) recorded at 175 Hz. (b)
Calculated Lorentzian lines for the electronic 'p'-state multiplet in
distorted and undistorted spherical nanocrystals. The corresponding
lifetimes were taken to be $100$ ps and $30$ ps. Note the splitting for
different values of $F$, which is weaker in the distorted nanocrystal.} \label{fig:1}
\end{center}
\end{figure}

The broad ODMR bands are reminiscent of cyclotron resonance data with
cyclotron effective mass values in good agreement with the well known
conduction and valence band effective masses of CdSe. It is evident,
however, that well defined cyclotron orbits cannot exist within the
nanocrystal's interior since the minimal cyclotron radius of an electron
under the magnetic field used in such experiments is much larger than the
size of a nanocrystal. Thus, the observed magnetic resonances cannot be
associated with the field induced diamagnetic currents within the
nanocrystals and should be assigned to a paramagnetic effect. They are most
probably associated with transitions between initial extended electron
(hole) states within the entire space of the nanocrystal and final localized
hole (electron) states trapped in defect cites (shown schematically in
Fig.2). The magnetic resonance events take place during the initial stages.

Focusing on the magnetic resonance processes we follow an approach exploited
in Ref. (~\cite{Gel'mont2}), where the g-factor for acceptor levels in
diamond-type semiconductors was calculated. The spherical symmetry of the
electron (hole) wavefunctions at the $\Gamma $-point of the Brillouin zone
simplifies the calculation considerably, leaving the crystal field and
lattice effects to appear only through effective masses and g-factors.
Within this approach the spherical symmetry is broken only at the
nanocrystal surface, namely when some of the nanocrystal faces do not follow
the elementary Wigner-Seitz polyhedron of the zinc-blende structure.
Electron and hole are treated throughout the paper as independent particles.
This is justified by the large electronic energy separation in a nanocrystal
with respect to the electron-hole Coulomb interaction (\cite{Efros1}).

Near the $\Gamma $-point the conduction band of bulk CdSe is nearly
parabolic, with isotropic effective mass tensor. Utilizing the effective
mass and g-factor of an electron provided by the ${\bf k}\cdot {\bf p}$
theory, the relevant Hamiltonian is written in the form, 
\begin{equation}
\hat{H}=\frac{1}{2m^{\ast }}\left( {\bf p}-\frac{e}{c}{\bf A}\right)
^{2}+g^{\ast }\mu _{B}{\bf S}\cdot {\bf B}+\hat{H}_{S.O.}  \label{1}
\end{equation}
where $m^{\ast }$ is the conduction band effective mass at the $\Gamma $
-point, $g^{\ast }$ is the corresponding effective g-factor (i.e. corrected
by the crystal field), and $\hat{H}_{S.O.}$ is the spin-orbit (SO)
interaction.

Selecting the symmetric gauge form of the vector potential ${\bf A}=(0,0,%
\frac{1}{2}Brsin\theta )$ to describe a constant and uniform magnetic field $%
{\bf B}$ aligned along the $z$-axis, with spherical polar coordinates $%
(r,\theta ,\phi )$ , Eq.(\ref{1}) can be rewritten in the form: $\hat{H}=%
{\bf p}^{2}/2m^{\ast }+\hat{H}_{S.O.}+\beta (L_{z}+g_{S}S_{z})B$, where the
diamagnetic term, $(eBrsin\theta )^{2}/8m^{\ast }c^{2}$, is neglected. Here $%
\beta =\mu _{B}m_{0}/m^{\ast }$, $g_{S}=g^{\ast }m^{\ast }/m_{0}$, $m_{0}$
is the free electron mass, and $L_{z}$ and $S_{z}$ are, respectively, the
projections of the orbital and spin angular momentum on the z-axis. In the
presence of the spin-orbit interaction the eigenstates correspond to the
total angular momentum ${\bf F}={\bf L}+{\bf S}$ , and the corrections to
the electron energy due to the paramagnetic term are $\delta
E_{el}=g_{F,el}\beta M_{F}B$, where $g_{F,el}$ is the electronic Lande
factor, and $M_{F}$ is the projection of ${\bf F}$ on the z-axis. Using a
standard scheme, we find that $g_{F}=1+(g_{S}-1)\alpha $, where $\alpha =%
\left[ S(S+1)+F(F+1)-L(L+1)\right] /2F(F+1)$.

The corresponding calculation near the valence band edge is considerably
more complicated due to the four-fold degeneracy of the band edge at the $%
\Gamma $-point. The relevant magnetic resonance transitions involve the
Zeeman splitting of a hole near this energy. Since the energies involved are
much smaller than the SO split-off energy $\Delta $, the latter can be
ignored \ (\cite{Gel'mont2}). Thus, the motion of holes near the band edge
in the presence of the magnetic field can be described by Luttinger
Hamiltonian (\cite{Luttinger2}),

\begin{equation}
\hat{D}=\frac{1}{m_{0}}\left\{ (\gamma _{1}+\frac{5}{2}\gamma)\frac{k^{2}}{2}%
-\gamma ({\bf k}\cdot {\bf J})^{2}+(\kappa-\frac{\gamma }{2})\frac{e}{c}{\bf %
J}\cdot {\bf B}\right\} ,  \label{5}
\end{equation}
where $\gamma $, $\gamma _{1}$, and $\kappa $ are Luttinger band parameters, 
${\bf k}={\bf p}-\frac{e}{c}{\bf A}$ and ${\bf J}$ is the set of $4\times 4$
matrices of spin $3/2$. In the absence of the magnetic field the ${\bf k}%
\cdot {\bf p}$ Hamiltonian at the $\Gamma$-point is invariant under
rotations so that the total angular momentum ${\bf F}={\bf L}+{\bf J}$ is
conserved, while the resulting energy eigenvalues are independent of $M_{F}$
. The perturbation theory with respect to the paramagnetic term yields for
the Zeeman splitting of a hole near the band edge $\delta E_{h}=g_{F,h}\mu
_{B}M_{F}B$, where $g_{F,h}$ is the Lande factor of a hole.

\begin{figure*}[htbp]
\begin{center}
\psfig{file=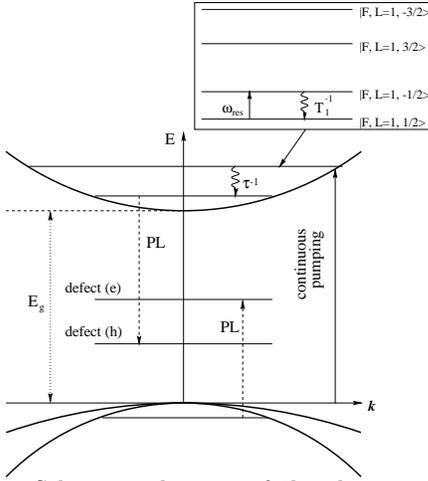,width=2.5in,angle=-90}
\caption{ Schematic drawing of the electronic spin-orbit energy levels
invloved in the proposed ODMR mechanism. The spin-orbit 'p'-state manifold
is shown in the inset. The relevant time scales are the Spin-lattice
relaxation time $T_{1}\sim 1-100$ $\mu $ s. , the non-radiative life-time $%
\tau _{p}\sim 1-200$ ps, and the magnetic resonance period $\omega
_{res}^{-1}\sim 100$ ps.} \label{fig:2}
\end{center}
\end{figure*}

The calculation of $g_{F,h}$ in the general case is complicated by the fact
that in the corresponding unperturbed wavefunctions heavy and light hole
components are mixed. A good approximation of $g_{F,h}$ can be obtained,
however, in the limit $F\gg 1$ , when the hole radial wavefunctions $%
R_{F,l}\left( r\right) $ (\cite{Grigoryan}) can be approximated by their
asymptotic forms for which the motion of the heavy and light holes are
completely decoupled. Consequently, the Zeeman energies of the holes can be
written in a separable form, $\delta E^{hh}=[\gamma _{1}-\frac{1}{2}\gamma
]\mu _{B}M_{F}B,\delta E^{lh}=[\gamma _{1}+\frac{1}{2}\gamma ]\mu
_{B}M_{F}B, $ where 'lh' and 'hh' stand for light and heavy hole,
respectively.\ \ The relevance of these simple limiting expressions to the
present study will be discussed below.

The above analysis relied on the spherical symmetry of the nanocrystals
under study. TEM pictures (\cite{Kadavanich}) however, showed that the
nanocrystals are slightly elongated spheroids or ellipsoids. The effect of
this symmetry breaking on the relevant electron and hole paramagnetic
states, was estimated by assuming small deviations from the spherical shape
and by exploiting the perturbation theory. A uniaxial shape distortion
introduces an energy, $\hat{H}_{SD}$, which is typically much smaller than $%
\hat{H}_{SO}$ , but $\ $much larger than the Zeeman spitting$\ \hat{H}_{Z}$
(see below). Thus, the perturbative procedure can be carried out in two
stages, within the framework of the distorted wave Born approximation.

We consider the case of an electron near the bottom of the conduction band
and use a model of a free particle enclosed in a nearly spherical cavity
slightly distorted along some axis. The first order correction to energy due
to this distortion is given by $\delta E_{n,l,F,M_{F}}^{SD}=\epsilon
E_{n,l,F}^{0}\left[ M_{F}^{2}/F(F+1)-1/3\right] $, where $E_{n,l,F}^{0}$ is
the corresponding unperturbed energy, $\epsilon =2(a-b)/(a+b)$ is the
deformation parameter, assumed to be small, $a$ and $b$ are the principal
semiaxes of the ellipsoid. Note that in the latter formula the z-axis is
selected along the symmetry axis of the nanocrystal.

Due to this uniaxial distortion the original $2F+1$ fold degenerate
multiplet splits into a collection of Kramers' doublets, each of which
having the same value of $|M_{F}|$. For the prolate nanocrystals the lowest
lying doublet corresponds to $M_{F}=\pm 1/2$ and the highest one has $%
|M_{F}|=F$. In the oblate case the picture is reversed.

It is important to note that, due to the finite size of the electron
wavefunctions ( i.e. $R\sim 3-4nm$ ), the minimal value of the unperturbed
energy $E_{n,l,F}^{0}\sim \hbar ^{2}R^{-2}/2m^{\ast }$ is typically much
larger than the Zeeman energy $\sim \beta M_{F}B$; consequently, the
distortion energy, $\delta E_{n,l,F,M_{F}}^{SD}$ , is significantly larger
than $\beta M_{F}B$ .

An external static magnetic field ${\bf B}$, lifts the residual degeneracy
of the Kramers' doublets. When the field is not aligned along the
nanocrystal's symmetry axis, one should use the more general form $\hat{H}%
_{Z}=g_{F}\beta {\bf F}\cdot {\bf B}$. The corresponding Zeeman splitting
energy of the doublets with $|M_{F}|=1/2$ is not given by the familiar
formula $\delta E_{Z}=g_{F}\mu _{B}Bcos\theta _{0}M_{F}$, where $\theta _{0}$
is the angle between ${\bf B}$ and the symmetry axis of the ellipsoid, since 
$\hat{H}_{Z}$ mixes the zero-order functions of the doublet with $M_{F}=\pm
1/2$. A slightly more complicated expression, $\delta E_{Z}=\pm 1/2g_{F}\mu
_{B}B[cos^{2}\theta _{0}+(F+1/2)^{2}sin^{2}\theta _{0}]^{1/2}$ , is obtained
in this case, which reduces however to the $\theta _{0}=0$ expression, $%
\delta E_{el}$ , at $F=1/2$ for any $\theta _{0}$. This dependence of the
Zeeman splitting energy on $F$ and $\theta _{0}$ requires a careful
calaculation of the observable position of the magnetic resonance peak,
which takes into account the transition matrix element $M_{-1/2%
\leftrightarrow 1/2}$. It is easy to show that $\left|
M_{-1/2\leftrightarrow 1/2}\right| =A(F+\frac{1}{2})\left| \cos 2\theta
_{0}\right| /2\left[ \cos ^{2}\theta _{0}+(F+\frac{1}{2})^{2}\sin ^{2}\theta
_{0}\right] ^{1/2}$ , where $A$ is constant, independent of $\theta _{0}$
and $F$.

Application of the microwave magnetic field in a plane perpendicular to $%
{\bf B}$ causes transitions of significant intensity between different
Zeeman split sublevels satisfying the selection rule $\Delta M_{F}=\pm 1$.
Considering first only the nanocrystals with symmetry axes parallel to ${\bf %
B}$ ( i.e. where $\theta _{0}=0$ ), it can be seen that among all the
transitions satisfying the selection rule, only those originating from the $%
M_{F}=\pm 1/2$ doublet can be observed at magnetic field $B$ in the
experimentally accessible range. This is due, on the one hand, to the large
distortion energy $\delta E_{n,l,F,M_{F}}^{SD}$ and, on the other hand, to
the the Kramers' degeneracy of the $M_{F}=\pm 1/2$ states in the distorted
nanocrystals. Similar behavior is anticipated for the hole transitions,
though the corresponding quantitative analysis of the shape distortion is
quite involved (\cite{Efros2}).

The present ODMR spectra was monitored at the PL broad band, centered well
below the bandgap energy. Thus it is assigned to transitions between excited
electrons (holes) within spin-orbit states at the conduction (valence) band,
recombining with holes (electrons) in defect levels, as shown schematically
in Fig.2. As mentioned earlier, the spin-lattice relaxation time, $T_{1}$ ,
of the excited electrons and holes is in the range of $10^{-4}$sec, while
that of the trapped defect is of two orders of magnitude larger, thus
suggesting that the spin flipping at the defect site can be considered
stationary at the time scale of the spin flip within the spin-orbit
manifold. Therefore, the spin splitting of the defect states was ignored in
the model shown in Fig.2.

Let us first consider the ODMR band associated with the conduction
electrons. For the excitation energy of the continuous $Ar^{+}$ laser used
in the experiment only the ground ('s'-state; $S=1/2,L=0,F=1/2$) and the
first few excited $F$-states can be populated. Although the electrons pumped
into an excited state relax quickly to the ground state via nonradiative
channels, the observed ODMR band may be considered as arising mainly from $%
M_{F}=1/2\rightarrow -1/2$ transitions in such an excited state. Indeed,
recent experiments have shown that in CdSe nanocrystals the life time, $\tau
_{p}$ , of the first excited state ('p'-state; $S=1/2,L=1,F=1/2$ and $%
S=1/2,L=1,F=3/2$) is in the range of $1-200\times 10^{-12}$sec. (\cite
{Klimov}, \cite{Guyot-Sionnest}). The corresponding decay rate, $\tau
_{p}^{-1}$,which is much larger than the spin relaxation rate, $T_{1}^{-1}$
, is of the same order of magnitude as the resonance microwave frequency, $%
\omega _{res}$ (see Fig.2). These imply that the spin excitation of
electrons occupying the 'p'-state by the mw source can be completed before
relaxing into the long-lived ( luminescent ) 's'-state, where their spin
polarization is preserved for time much longer than the characteristic
recombination time.

Assuming Lorentzian densities of $F$-states centered at $\delta E_{Z}$ with
frequency spread corresponding to $\tau _{p}^{-1}$ , and using the above
expression for $M_{-1/2\leftrightarrow 1/2}$ , we find , after averaging
over the angle $\theta _{0}$ , the resonance lines shown in Fig.(1b) for
typical values of $\tau _{p}$. For\ the selected well known value of the
electron effective mass $m^{\ast }=0.13m_{0}$ , the calculated resonance
position is in a very good agreement with the lower field resonance band
observed in the experimental ODMR spectrum. In contrast , the calculated
resonance peak due to spin excitations in the 's'-state appears at a field
about 1.1T , well out of the field range shown in Fig.(1). It is interesting
to note that the calculated integral intensity of this resonance is found to
be much weaker than that of the 'p' state described above.

Evaluation of $g_{F,h}$ for the low-laying hole states has proved to be
quite difficult. The corresponding resonance peak positions can be
estimated, however, if we note that for $g_{F,el}$ in spherical nanocrystal
all the levels of the multiplet with $L>0$ ( i.e. with the exception of $%
L=0, $ $F=1/2$ ) cluster in the close vicinity of the multiplet's centre.
The latter can be found by taking the mathematical limit $F\rightarrow
\infty $ \ in the formula for $\delta E_{el}$. Assuming that the resonance
lines of holes follow a similar pattern and using the above formulae for $%
\delta E^{hh}$ and $\delta E^{lh}$ together with the valence band parameters
reported in ~( \cite{Norris}) the centres of the respective multiplets are
located at $B_{hh}=h\nu /(\gamma _{1}-\frac{1}{2}\gamma )\mu _{B}$, $%
B_{lh}=h\nu /(\gamma _{1}+\frac{1}{2}\gamma )\mu _{B}$, yielding $B_{hh}$ $%
=0.44$ Tesla and $B_{lh}$ $=0.33$ Tesla respectively. The value of $B_{lh}$
is in close agreement with the maximum of the middle broad band shown in
Fig.1. However, $B_{hh}$ deviates slightly from the highest field band
maximum, which is not surprising in view of the approximate nature of the
present calculation.

In conclusion it was shown that the proposed mechanism of electron (hole)
paramagnetic resonance in nearly spherical zinc-blende semiconductor
nanocrystals accounts well for the unusual features observed in the ODMR
spectra of $CdSe$ nanoparticles. The observability of the corresponding
resonance bands is due to the small size (i.e. in the range of several
nanometers) of the crystallites as well as to the Kramers' degeneracy of the
energy levels, which cannot be lifted by any shape distortion of
electrostatic origin.

The authors express their deepest gratitude to Prof. G. Kventsel, V.
Zhuravlev and R. Guliamov for useful discussions. This research was
supported by the Israel Science Foundation, grant no. 36599-12.5

\end{document}